\documentclass[12pt]{article}
\usepackage{amssymb,amsmath,comment,mathtools}
\usepackage[dvipdfmx]{graphicx}
\usepackage{subfigmat}
\usepackage{braket}
\usepackage{hyperref} 	  		
\usepackage{xcolor}
\usepackage{physics}
\usepackage{booktabs}
\usepackage{multirow}
\usepackage{array}
\usepackage{epsfig}
\usepackage{here}
\usepackage{color}
\usepackage{bm}
\graphicspath{{./Figures/}}

\newcommand{\ba}{\begin{align}}
\newcommand{\ea}{\end{align}}

\def\bea{\begin{eqnarray}}
\def\eea{\end{eqnarray}}
 \makeatletter
\def\alt{\mathrel{\mathpalette\gl@align<}}
\def\agt{\mathrel{\mathpalette\gl@align>}}
\def\gl@align#1#2{\lower.6ex\vbox{\baselineskip\z@skip\lineskip\z@
\ialign{$\m@th#1\hfil##\hfil$\crcr#2\crcr\sim\crcr}}} \makeatother
\renewcommand{\thefootnote}{\fnsymbol{footnote}}

\begin{document}
\begin{flushright}
\end{flushright}
\vspace*{1.0cm}

\begin{center}
\baselineskip 20pt 
{\Large\bf 
Gravitational Waves from Phase Transition in a Supersymmetric Left-Right Model
}
\vspace{1cm}

{\large 
Naoyuki Haba${}^{a,b}$,  Yasuhiro Shimizu${}^{a,b}$ \\
and Toshifumi Yamada${}^{c}$
} \vspace{.5cm}

{\baselineskip 20pt \it
${}^{a}$Department of Physics, Osaka Metropolitan University, Osaka 558-8585, Japan \\
${}^{b}$Nambu Yoichiro Institute of Theoretical and Experimental Physics (NITEP),
Osaka Metropolitan University, Osaka 558-8585, Japan\\
${}^{c}$Institute for Mathematical Informatics, Meiji Gakuin University, Yokohama 244-8539, Japan
}

\vspace{.5cm}

\vspace{1.5cm} {\bf Abstract} \end{center}

We investigate the cosmological phase transition dynamics in a supersymmetric left-right symmetric model based on the gauge group $SU(3)_C \times SU(2)_L \times SU(2)_R \times U(1)_{B-L}$ that addresses the strong CP problem through extended parity symmetry and doublet-doublet splitting. We compute the finite temperature effective potential including one-loop Coleman-Weinberg corrections, thermal contributions, and daisy resummation to determine whether the $SU(2)_R \times U(1)_{B-L} \to U(1)_Y$ symmetry breaking transition can produce observable gravitational waves. For phenomenologically viable parameters satisfying current LHC constraints, we find that the phase transition is strongly first-order with nucleation temperature $T_n \sim 0.5 v_R$, transition strength parameter $\alpha \sim 0.01-0.3$, and inverse duration $\beta/H \sim 100$. The resulting stochastic gravitational wave background peaks at frequencies $f \sim 0.1-1$ Hz with amplitude $h^2\Omega_{GW} \sim 10^{-14}-10^{-12}$. We find that there is a parameter region where the gravitational wave spectrum overlaps with DECIGO/BBO sensitivity curves, providing a potentially observable signature connecting the theoretical solution to the strong CP problem with gravitational wave experiments.

\thispagestyle{empty}

%\bigskip
\newpage
\renewcommand{\thefootnote}{\arabic{footnote}}
%\addtocounter{page}{-1}
\setcounter{footnote}{0}
%%%%%%%%%%%%%%%%%%%%%%%%%%
%\baselineskip 36pt
% Main body
%%%%%%%%%%%%%%%%%%%%%%%%%%
\baselineskip 18pt
%%%%%%%%%%%%%%%%%%%%%%%%%%

\title{Gravitational Waves from Phase Transition in a Supersymmetric Left-Right Model}
\author{Naoyuki Haba${}^{a,b}$, Yasuhiro Shimizu${}^{a,b}$ and Toshifumi Yamada${}^{c}$\\[0.5em]
${}^a$Department of Physics, Osaka Metropolitan University, Osaka, Japan\\
${}^b$Nambu Yoichiro Institute of Theoretical and Experimental Physics (NITEP), \\
Osaka Metropolitan University, Osaka 558-8585, Japan\\
${}^c$Institute for Mathematical Informatics, Meiji Gakuin University,\\
Yokohama 244-8539, Japan}

\section{Introduction}

The Standard Model (SM) of particle physics, while remarkably successful in describing experimental observations, leaves several fundamental questions unanswered. Among these, the strong CP problem is one of the most persistent theoretical puzzles. Several theoretical approaches have been proposed to address this problem. The most widely studied is the Peccei-Quinn mechanism~\cite{Peccei1977}, which introduces a new global $U(1)$ symmetry that is spontaneously broken, leading to a pseudo-Nambu-Goldstone boson called the axion~\cite{Kim1979, Shifman1980, Dine1981, Zhitnitsky1980}. However, concerns about explicit breaking of the Peccei-Quinn symmetry by quantum gravity effects have motivated alternative approaches.

Left-right symmetric models provide an attractive alternative framework for addressing the strong CP problem~\cite{Kuchimanchi1996, Mohapatra1996}. These models restore parity symmetry at high energies, naturally explaining the absence of strong CP violation. The basic idea is that if nature is parity-symmetric at high energies, then CP-violating phases can be rotated away in the quark mass matrices, solving the strong CP problem. Early proposals, however, faced the difficulty that CP-violating phases could be regenerated through renormalization group evolution when the left-right symmetry is broken, potentially reintroducing the strong CP problem at low energies. Supersymmetric versions of left-right models have been extensively studied~\cite{Aulakh:1997fq, Aulakh1999, Aulakh2000}.  However, these early models still suffered from the problem of CP violation regeneration and faced difficulties in achieving realistic fermion masses and mixings while maintaining gauge coupling unification.

We consider a novel supersymmetric left-right symmetric model that overcomes these limitations through a "doublet-doublet splitting" mechanism. Unlike previous models that break the left-right symmetry using $SU(2)_R$ triplet fields~\cite{Aulakh:1997fq}, this approach employs $SU(2)_R$ doublets for symmetry breaking. The key innovation is that when the $SU(2)_R$ symmetry is broken, only one combination of the $SU(2)_L$ Higgs doublets remains light, effectively reducing the low-energy theory to the Minimal Supersymmetric Standard Model (MSSM). This mechanism prevents the regeneration of CP-violating phases until the six-loop level, providing a robust solution to the strong CP problem.

The cosmological implications of such models are equally intriguing. Phase transitions in the early universe associated with spontaneous symmetry breaking can generate stochastic backgrounds of gravitational waves through bubble nucleation, collision, and sound wave processes. The detection of gravitational waves by LIGO~\cite{LIGOScientific:2016aoc} has opened a new window for probing fundamental physics, and future space-based detectors such as LISA~\cite{LISA2017}, BBO~\cite{Crowder2005}, and DECIGO~\cite{Seto2001, Kawamura2006} will be sensitive to gravitational waves from cosmological phase transitions.

The purpose of this paper is to analyze the phase transition dynamics in this model and to assess its potential for producing observable gravitational wave signatures. We compute the finite temperature effective potential for the model, including one-loop Coleman-Weinberg corrections, thermal contributions, and daisy resummation to handle infrared divergences that arise at finite temperature. We determine the critical parameters characterizing the phase transition—the critical temperature $T_c$, nucleation temperature $T_n$, strength parameter $\alpha$, and inverse duration $\beta/H$—and calculate the resulting gravitational wave spectrum using state-of-the-art methods developed for cosmological phase transitions.

Our analysis reveals that the $SU(2)_R \times U(1)_{B-L} \to U(1)_Y$ phase transition is indeed strongly first-order for phenomenologically viable parameter choices, with the ratio $v_c/T_c > 1$ required for efficient gravitational wave production. The resulting gravitational wave spectrum has a peak frequency in the range $f \sim 10^{-1} - 1$ Hz. We find that there is a parameter region where the gravitational wave spectrum overlaps with the DECIGO/BBO sensitivity curves, providing a potentially observable signature connecting the theoretical solution to the strong CP problem with gravitational wave experiments.

\section{Supersymmetric Left-Right Model}

In this section, we provide a detailed description of the supersymmetric left-right symmetric model. We begin with the gauge structure and field content, followed by the superpotential and symmetry breaking pattern that are crucial for understanding both the solution to the strong CP problem and the phase transition dynamics.

\subsection{Gauge Group and Symmetries}

The model is based on the gauge group:
\begin{equation}
G = SU(3)_C \times SU(2)_L \times SU(2)_R \times U(1)_{B-L}
\end{equation}

This extends the Standard Model gauge group by promoting the right-handed fermions to transform under an $SU(2)_R$ gauge symmetry and introducing a $U(1)_{B-L}$ factor that distinguishes baryon number minus lepton number. The left-right symmetry is manifest in the symmetric treatment of left and right-handed fermions under their respective $SU(2)$ groups.

Beyond the gauge symmetries, the model incorporates two additional discrete symmetries:

\subsubsection{Extended Parity Symmetry}

The model includes an extended parity transformation $\mathcal{P}$ that exchanges left and right sectors:
\begin{align}
\mathcal{P}: \quad SU(2)_L &\leftrightarrow SU(2)_R \\
Q_L &\leftrightarrow Q_R^c \\
L_L &\leftrightarrow L_R^c
\end{align}

This parity symmetry ensures that the Yukawa coupling matrices are Hermitian in flavor space, which is a key ingredient in eliminating CP violation from the strong sector.

\subsubsection{Discrete R-Symmetry}

The model also possesses a discrete $R$-symmetry $Z_{2n}^R$ where $n$ is an odd integer with $n \geq 3$. This $R$-symmetry plays crucial roles in controlling the superpotential structure, ensuring vacuum stability, and preventing unwanted terms that could reintroduce CP violation. The discrete nature allows for small controlled violations necessary for realistic phenomenology while maintaining essential features for solving the strong CP problem.

\subsection{Field Content and Quantum Numbers}

The model contains an extended spectrum of chiral superfields. The complete field content is presented in Table~\ref{tab:fields}.

\begin{table}[h]
\centering
\begin{tabular}{|c|c|c|c|c|c|c|}
\hline
\multirow{2}{*}{Superfield} & \multicolumn{4}{|c|}{Gauge Quantum Numbers} & \multirow{2}{*}{$Z_{2nR}^R$} & \multirow{2}{*}{Generation} \\
\cline{2-5}
 & $SU(3)_C$ & $SU(2)_L$ & $SU(2)_R$ & $U(1)_{B-L}$ & & \\
\hline
\hline
\multicolumn{7}{|c|}{Matter Fields} \\
\hline
$Q^i$ & $\mathbf{3}$ & $\mathbf{2}$ & $\mathbf{1}$ & $1/3$ & $n$ & $i=1,2,3$ \\
$Q^{ci}$ & $\overline{\mathbf{3}}$ & $\mathbf{1}$ & $\mathbf{2}$ & $-1/3$ & $n$ & $i=1,2,3$ \\
$L^i$ & $\mathbf{1}$ & $\mathbf{2}$ & $\mathbf{1}$ & $-1$ & $n$ & $i=1,2,3$ \\
$L^{ci}$ & $\mathbf{1}$ & $\mathbf{1}$ & $\mathbf{2}$ & $1$ & $n$ & $i=1,2,3$ \\
$N^i$ & $\mathbf{1}$ & $\mathbf{1}$ & $\mathbf{1}$ & $0$ & $2-n$ & $i=1,2,3$ \\
\hline
\multicolumn{7}{|c|}{Higgs Fields} \\
\hline
$\Phi_a$ & $\mathbf{1}$ & $\mathbf{2}$ & $\mathbf{2}$ & $0$ & $2$ & $a=1,2$ \\
$H$ & $\mathbf{1}$ & $\mathbf{2}$ & $\mathbf{1}$ & $-1$ & $0$ & -- \\
$\bar{H}$ & $\mathbf{1}$ & $\mathbf{2}$ & $\mathbf{1}$ & $1$ & $0$ & -- \\
$H^c$ & $\mathbf{1}$ & $\mathbf{1}$ & $\mathbf{2}$ & $1$ & $0$ & -- \\
$\bar{H}^c$ & $\mathbf{1}$ & $\mathbf{1}$ & $\mathbf{2}$ & $-1$ & $0$ & -- \\
$S$ & $\mathbf{1}$ & $\mathbf{1}$ & $\mathbf{1}$ & $0$ & $2$ & -- \\
\hline
\end{tabular}
\caption{Field content of the supersymmetric left-right model.}
\label{tab:fields}
\end{table}

\subsubsection{Matter Fields}

The matter sector contains three generations organized to respect left-right symmetry:
\begin{itemize}
\item $Q^i = (u_L^i, d_L^i)^T$: Left-handed quark doublets
\item $Q^{ci} = (u_R^{ci}, d_R^{ci})^T$: Right-handed quark doublets under $SU(2)_R$
\item $L^i = (\nu_L^i, e_L^i)^T$: Left-handed lepton doublets  
\item $L^{ci} = (\nu_R^{ci}, e_R^{ci})^T$: Right-handed lepton doublets under $SU(2)_R$
\item $N^i$: Gauge singlet sterile neutrinos for the inverse seesaw mechanism
\end{itemize}

\subsubsection{Higgs Fields}

The Higgs sector is considerably richer than the MSSM:
\begin{itemize}
\item $\Phi_a$ ($a = 1,2$): Bi-fundamental fields $(\mathbf{2}, \mathbf{2})$ under $SU(2)_L \times SU(2)_R$ that mediate between left and right sectors and implement doublet-doublet splitting
\item $H, \bar{H}$: $SU(2)_L$ doublets with $U(1)_{B-L}$ charges $\mp 1$
\item $H^c, \bar{H}^c$: $SU(2)_R$ doublets with $U(1)_{B-L}$ charges $\pm 1$ responsible for breaking $SU(2)_R \times U(1)_{B-L}$ symmetry
\item $S$: Gauge singlet field driving symmetry breaking dynamics
\end{itemize}

\subsection{Superpotential Structure}

The superpotential can be constructed to respect the symmetry structure while allowing realistic phenomenology. It decomposes into $R$-conserving and $R$-violating parts.

The R-symmetric part contains the dominant interactions:
\begin{align}
W_R &= \sum_{i,a} Y_{qa}^{ij} Q^i \Phi_a Q^{cj} + \sum_{i,a} Y_{\ell a}^{ij} L^i \Phi_a L^{cj} \nonumber \\
&\quad + \sum_i Y_n^{ij} N^i \bar{H} L^j + \sum_i Y_n^{ij} N^i \bar{H}^c L^{cj} \nonumber \\
&\quad + \sum_a \kappa_a H \Phi_a H^c + \sum_a \kappa'_a \bar{H} \Phi_a \bar{H}^c \nonumber \\
&\quad + S\left(\lambda^* \bar{H}H + \lambda \bar{H}^c H^c - \xi\right)
\end{align}

The Yukawa coupling matrices $Y_{qa}^{ij}$ and $Y_{\ell a}^{ij}$ are constrained by extended parity to be Hermitian: $Y_{qa}^{ij} = (Y_{qa}^{ji})^*$, which is crucial for CP conservation. The neutrino terms implement the inverse seesaw mechanism, while the mixing terms $\kappa_a, \kappa'_a$ are the heart of the doublet-doublet splitting mechanism. The singlet $S$ drives symmetry breaking with parameter $\xi$ setting the scale.

Small R-symmetry breaking terms are essential for realistic phenomenology:
\begin{align}
W_{R\!\!\!/} &= \frac{1}{2}\sum_{a,b} \mu_{ab} \Phi_a \Phi_b + \frac{1}{2}\sum_i \mu_N (N^i)^2 \nonumber \\
&\quad + m_H^* \bar{H}H + m_H \bar{H}^c H^c + \frac{1}{2}m S^2 + \frac{1}{3}k S^3
\end{align}

These terms satisfy the hierarchy condition $|\mu_{ab}|, |\mu_N|, |m_H|, |m| \ll \sqrt{|\xi|}$, ensuring proper doublet-doublet splitting while providing controlled R-symmetry breaking.

\subsection{Symmetry Breaking Pattern and Vacuum Structure}

The model undergoes symmetry breaking in two stages:

\begin{equation}
SU(3)_C \times SU(2)_L \times SU(2)_R \times U(1)_{B-L} \xrightarrow{\langle H^c \rangle, \langle \bar{H}^c \rangle} SU(3)_C \times SU(2)_L \times U(1)_Y
\end{equation}

This occurs at scale $v_R \sim 10^{5-7}$ GeV with VEV pattern:
\begin{align}
\langle H^c \rangle &= \begin{pmatrix} 0 \\ v_R \end{pmatrix}, \quad \langle \bar{H}^c \rangle = \begin{pmatrix} v_R \\ 0 \end{pmatrix}
\end{align}

The hypercharge generator is $Y = T_R^3 + (B-L)/2$. 
At the electroweak scale when the light $SU(2)_L$ doublet acquires a VEV.
\begin{equation}
SU(3)_C \times SU(2)_L \times U(1)_Y \xrightarrow{\langle H \rangle} SU(3)_C \times U(1)_{em}
\end{equation}

\subsection{Solution to the Strong CP Problem}

The model provides a complete solution through the combination of left-right symmetry, supersymmetry, and doublet-doublet splitting.

Extended parity constrains Yukawa matrices to be Hermitian: $Y_{qa}^{ij} = (Y_{qa}^{ji})^*$. Combined with real Higgs VEVs from supersymmetry, this yields Hermitian quark mass matrices:
\begin{equation}
M_q = \sum_a Y_{qa} v_{\Phi_a}
\end{equation}

Since Hermitian mass matrices can be diagonalized by unitary transformations preserving CP, there are no physical CP-violating phases in the quark sector.

After $SU(2)_R$ breaking, the mixing terms generate a mass matrix for $SU(2)_L$ doublets:
\begin{equation}
\mathcal{M}_{LL}^2 = \begin{pmatrix}
|\kappa_a|^2 v_R^2 + m_H^2 & |\kappa_a \kappa'_a| v_R^2 + m_H^{*2} \\
|\kappa_a \kappa'_a| v_R^2 + m_H^2 & |\kappa'_a|^2 v_R^2 + m_H^{*2}
\end{pmatrix}
\end{equation}

This has eigenvalues of order $v_R^2$ (heavy) and $m_H^2 \ll v_R^2$ (light). Only the light combination survives at low energies, giving an effective MSSM where CP violation can only enter at six-loop level—a negligible effect that maintains the solution to the strong CP problem.

\subsection{Experimental Constraints on the Symmetry Breaking Scale}

The symmetry breaking scale $v_R$ associated with $SU(2)_R \times U(1)_{B-L}$ breaking is subject to stringent experimental constraints from collider searches and precision measurements. Direct searches at the LHC for right-handed gauge bosons $W_R$ decaying to lepton-neutrino pairs provide the most direct constraints on $v_R$. The recent searches from the CMS Collaboration, using the full Run-2 dataset of 138 fb$^{-1}$ at $\sqrt{s} = 13$ TeV, exclude $W_R$ masses below approximately $4.7$--$5.4$ TeV at 95\% confidence level, depending on the right-handed neutrino mass spectrum and the decay channel~\cite{CMS:2021mjp}. Similarly, ATLAS analyses with 139 fb$^{-1}$ exclude $W_R$ (or $W'$) bosons with Standard Model-like couplings below $5.1$--$6.0$ TeV~\cite{ATLAS:2019lsy}. For gauge coupling values in the range $g_R \sim 0.6$--$0.8$ relevant to our analysis, these limits translate to a lower bound $v_R \gtrsim (5-8) \times 10^{3}$ GeV. Additionally, electroweak precision observables constrain the mixing between left- and right-handed gauge bosons, which in our model is suppressed by the ratio $v^2/v_R^2$ where $v = 246$ GeV is the electroweak scale. The stringent agreement between Standard Model predictions and LEP/SLD measurements requires $v_R \gtrsim 10^4$ GeV to maintain mixing angles below the percent level~\cite{Maiezza:2010ic}. 

\section{Finite Temperature Effective Potential}

To analyze the phase transition dynamics in the supersymmetric left-right model, we compute the complete finite temperature effective potential. This involves several components: the tree-level potential, one-loop Coleman-Weinberg corrections, thermal contributions, and daisy resummation to handle infrared divergences that arise at finite temperature. The calculation is crucial for determining whether the $SU(2)_R \times U(1)_{B-L}$ breaking transition is strongly first-order and capable of producing observable gravitational waves.

\subsection{General Framework and Field Parameterization}

The finite temperature effective potential can be written as \cite{Dolan1974} :
\begin{equation}
V_{\text{eff}}(\phi, T) = V_0(\phi) + V_{CW}(\phi) + V_T(\phi, T) + V_{\text{daisy}}(\phi, T)
\label{eq:total_potential}
\end{equation}

where $V_0(\phi)$ is the tree-level potential, $V_{CW}(\phi)$ is the one-loop Coleman-Weinberg potential, $V_T(\phi, T)$ represents thermal corrections, and $V_{\text{daisy}}(\phi, T)$ accounts for resummation of leading infrared divergences
\cite{Carrington1992}\cite{Curtin:2016urg}.

We focus on the phase transition associated with the breaking of $SU(2)_R \times U(1)_{B-L}$ symmetry, parameterized by the vacuum expectation values of the right-handed Higgs doublets. The field configuration during the phase transition is:
\begin{align}
\langle H^c \rangle &= \frac{1}{\sqrt{2}}\begin{pmatrix} 0 \\ v_R \end{pmatrix} \\
\langle \bar{H}^c \rangle &= \frac{1}{\sqrt{2}}\begin{pmatrix} v_R \\ 0 \end{pmatrix}
\end{align}

where $v_R$ is a real, positive field-dependent parameter characterizing the order parameter. We assume all other fields have vanishing VEVs during the high-temperature phase: $\langle H \rangle = \langle \bar{H} \rangle = \langle S \rangle = \langle N^i \rangle = 0$.

\subsection{Mass Spectrum and Field-Dependent Masses}

When the right-handed Higgs doublets acquire VEVs, various fields become massive through the Higgs and gauge mechanisms. We need the complete mass spectrum as functions of $v_R$.

\subsubsection{Gauge Boson Masses}

The $SU(2)_R$ and $U(1)_{B-L}$ gauge bosons acquire masses:

\textbf{Charged gauge bosons:}
\begin{equation}
M_{W_R}^2(v_R) = \frac{g_R^2 v_R^2}{2}
\end{equation}

\textbf{Neutral gauge bosons:} The neutral bosons $W_R^3$ and $B'$ mix. In the basis $(W_R^3, B')$, the mass-squared matrix is:
\begin{equation}
\mathcal{M}_{neutral}^2 = \frac{v_R^2}{2}\begin{pmatrix}
g_R^2 & -g_R g_{B-L} \\
-g_R g_{B-L} & g_{B-L}^2
\end{pmatrix}
\end{equation}

The eigenvalues are:
\begin{align}
M_{Z'}^2(v_R) &= \frac{(g_R^2 + g_{B-L}^2) v_R^2}{2} \\
M_{\gamma'}^2(v_R) &= 0
\end{align}

The massless state corresponds to the unbroken $U(1)_Y$ hypercharge.

\subsubsection{Scalar Masses}

The scalar sector is complex due to the extended Higgs content.

\textbf{Right-handed Higgs doublets:} Decomposing the fields as:
\begin{align}
H^c &= \frac{1}{\sqrt{2}}\begin{pmatrix} \sqrt{2}h_u^c \\ v_R + h_d^c + ia_d^c \end{pmatrix} \\
\bar{H}^c &= \frac{1}{\sqrt{2}}\begin{pmatrix} v_R + \bar{h}_u^c + i\bar{a}_u^c \\ \sqrt{2}\bar{h}_d^c \end{pmatrix}
\end{align}

The CP-even neutral components $(h_d^c, \bar{h}_u^c)$ have mass-squared matrix:
\begin{equation}
\mathcal{M}_{CP-even}^2 = \begin{pmatrix}
\frac{3g_R^2 + g_{B-L}^2}{2}v_R^2 & -\frac{g_R^2 + g_{B-L}^2}{4}v_R^2 \\
-\frac{g_R^2 + g_{B-L}^2}{4}v_R^2 & \frac{g_R^2 + 3g_{B-L}^2}{2}v_R^2
\end{pmatrix}
\end{equation}

The CP-odd and charged components have masses:
\begin{align}
m_{CP-odd}^2(v_R) &= \frac{g_R^2 v_R^2}{2} \\
m_{\pm}^2(v_R) &= \frac{g_R^2 v_R^2}{2}
\end{align}

\textbf{Bi-fundamental scalars:} The $\Phi_a$ fields develop masses through $\mu_{ab}$ terms and gauge interactions:
\begin{align}
m_{\Phi_u}^2(v_R) &= |\mu_{ab}|^2 + \frac{g_R^2 v_R^2}{4} + |\kappa_a|^2 v_R^2 \\
m_{\Phi_d}^2(v_R) &= |\mu_{ab}|^2 + \frac{g_R^2 v_R^2}{4} + |\kappa'_a|^2 v_R^2
\end{align}

\textbf{Left-handed Higgs doublets:} Through mixing terms:
\begin{align}
m_H^2(v_R) &= |\kappa_a|^2 v_R^2 + |m_H|^2 \\
m_{\bar{H}}^2(v_R) &= |\kappa'_a|^2 v_R^2 + |m_H|^2
\end{align}

\subsubsection{Matter Scalar Masses}

\begin{align}
  m_{\tilde{N}^c}^2(v_R) &=  |Y_N v_R|^2
\end{align}

\subsubsection{Fermionic Masses}

Gauginos mix with higgsinos when Higgs fields acquire VEVs. The neutralino mass matrix in the basis $(\tilde{B}', \tilde{W}^3_R, \tilde{h}^c_d, \tilde{\bar{h}}^c_u, \tilde{s})$, the dominant $5 \times 5$ block is:
\begin{equation}
\mathcal{M}_N^{(5)} = \begin{pmatrix}
M_{\tilde{B}'} & 0 & -g_{B-L}v_R & g_{B-L}v_R & 0 \\
0 & M_{\tilde{W}_R} & -g_R v_R & -g_R v_R & 0 \\
-g_{B-L}v_R & -g_R v_R & 0 & -m_H & \lambda v_R \\
g_{B-L}v_R & -g_R v_R & -m_H & 0 & \lambda v_R \\
0 & 0 & \lambda v_R & \lambda v_R & m
\end{pmatrix}
\end{equation}

The chargino mass matrix in the basis $(\tilde{W}^+_R, \tilde{h}^c_u)$ yields:
\begin{equation}
\mathcal{M}_C = \begin{pmatrix}
M_{\tilde{W}_R} & \sqrt{2}g_R v_R \\
0 & m_H
\end{pmatrix}
\end{equation}

\paragraph{Right-handed Neutrino Masses:}
\begin{equation}
M_{N^c}^2 = |Y_n|^2 v_R^2
\end{equation}

\subsection{Finite temperature effective potential Potential}

The complete finite temperature effective potential is:
\begin{equation}
V_{\text{eff}}(v_R, T) = V_{\text{tree}}(v_R) + V_{1-\text{loop}}(v_R) + V_T(v_R, T) + V_{\text{daisy}}(v_R, T).
\end{equation}

The tree-level potential is givne by
\begin{align}
V_{\text{tree}}(v_R) &= \left|\frac{1}{2}\lambda v_R^2 - \xi + m\frac{m_H}{\lambda} - k\frac{m_H^2}{\lambda^2}\right|^2 
\end{align}

The Coleman-Weinberg potential accounts for quantum corrections as follows:
\begin{equation}
V_{CW}(v_R) = \frac{1}{64\pi^2}\sum_i (-1)^{2s_i} n_i m_i^4(v_R) \left[\ln\left(\frac{m_i^2(v_R)}{\mu^2}\right) - c_i\right]
\end{equation}
where $s_i$ is the spin, $n_i$ the degrees of freedom, $\mu$ the renormalization scale, and $c_i = 3/2$ for minimal subtraction.

The most important contributions come from:
\begin{itemize}
\item \textbf{Gauge bosons:} $W_R^\pm$ (6 dof), $Z'$ (2 dof) with masses $\propto g v_R$
\item \textbf{Heavy Higgs scalars:} Various components with masses $\propto g v_R$ 
\item \textbf{Heavy fermions:} Neutralinos and charginos with masses $\propto g v_R$
\item \textbf{Matter scalars:} Squarks and sleptons with $D$-term masses $\propto g^2 v_R^2$
\end{itemize}

For typical parameter values, this gives corrections of order:
\begin{equation}
V_{CW} \sim \frac{g^4 v_R^4}{64\pi^2} \ln\left(\frac{v_R^2}{\mu^2}\right)
\end{equation}

\subsection{Thermal Corrections}

At finite temperature, thermal fluctuations modify the effective potential. The thermal contribution is:
\begin{equation}
V_T(v_R, T) = \frac{T^4}{2\pi^2} \sum_i (-1)^{2s_i} n_i J_{s_i}\left(\frac{m_i^2(v_R)}{T^2}\right)
\end{equation}

where the thermal functions are:
\begin{align}
J_0(x) &= \int_0^{\infty} dy \, y^2 \ln(1 - e^{-\sqrt{y^2+x}}) \quad \text{(scalars)} \\
J_{1/2}(x) &= \int_0^{\infty} dy \, y^2 \ln(1 + e^{-\sqrt{y^2+x}}) \quad \text{(fermions)}
\end{align}

For high temperatures ($T^2 \gg m_i^2$), these expand as:
\begin{align}
J_0(x) &= -\frac{\pi^4}{45} + \frac{\pi^2 x}{12} - \frac{x^2}{24}\ln\left(\frac{x}{a_0}\right) + \mathcal{O}(x^3) \\
J_{1/2}(x) &= \frac{7\pi^4}{360} - \frac{\pi^2 x}{24} - \frac{x^2}{48}\ln\left(\frac{x}{a_{1/2}}\right) + \mathcal{O}(x^3)
\end{align}

\subsection{Daisy Resummation}

At finite temperature, infrared divergences arise from massless or light modes, requiring resummation of leading contributions through "daisy" or "ring" diagrams.

\subsubsection{Thermal Masses}

The thermal mass corrections for different field types are:

\textbf{Gauge bosons:}
\begin{align}
\Pi_{W_R}(T) &= \frac{g_R^2 T^2}{6}(2 + N_H) \\
\Pi_{Z'}(T) &= \frac{T^2}{6}\left(g_R^2 + g_{B-L}^2 \sum_i q_i^2 N_i\right)
\end{align}

where $N_H$ counts Higgs doublets and the sum includes all fields charged under $U(1)_{B-L}$.

\textbf{Scalar fields:} For a scalar $\phi$ with gauge and scalar couplings:
\begin{equation}
\Pi_\phi(T) = \frac{T^2}{12}\left(\sum_a C_\phi^a g_a^2 + \sum_{\chi} \lambda_{\phi\chi}^2\right)
\end{equation}

For the right-handed Higgs doublets:
\begin{equation}
\Pi_{H^c}(T) = \frac{T^2}{12}\left(\frac{3g_R^2}{4} + g_{B-L}^2 + |\lambda|^2 + \sum_a |\kappa_a|^2\right)
\end{equation}

\subsubsection{Resummed Effective Potential}

The daisy-resummed potential replaces thermal masses with effective masses:
\begin{equation}
V_{\text{daisy}}(v_R, T) = V_T(v_R, T)\Big|_{m_i^2 \to m_i^2 + \Pi_i(T)} - V_T(v_R, T)
\end{equation}

This can be written as:
\begin{equation}
V_{\text{daisy}}(v_R, T) = -\frac{T}{12\pi}\sum_i n_i \left[\left(m_i^2(v_R) + \Pi_i(T)\right)^{3/2} - \left(m_i^2(v_R)\right)^{3/2}\right]
\end{equation}

The resummation is essential for handling light modes that can become tachyonic near the phase transition, ensuring the potential remains well-defined.

\subsection{$SU(2)_R\times U(1)_{B-L}$ Breaking Phase Transition}

In this section, we analyze the dynamics of the first-order phase transition associated with the breaking of $SU(2)_R \times U(1)_{B-L}$ symmetry in the supersymmetric left-right model. We compute the O(3)-symmetric Euclidean action for bubble nucleation, determine the critical parameters characterizing the phase transition, and assess the conditions for successful percolation.

We calculate the O(3)-symmetric Euclidean action \cite{Linde:1980tt}\cite{Linde:1981zj},for the high-temperature phase transition from the metastable vacuum $(v_R = 0)$ to the absolute vacuum $(v_R \neq 0)$. The transition is described by the bounce solution $v_R(r)$ that interpolates between these vacua. To compute the $O(3)$-symmetric Euclidean action, $S_E$ we use CosmoTransitions \cite{Wainwright:2011kj}. We verify numerically that for all parameter points studied, the vacuum with $\langle H^c \rangle, \langle \bar{H}^c \rangle \neq 0$ is the global minimum at T=0, ensuring cosmological stability.

Using the action, $S_E$, we can calculate the nucleation
temperature, $T_n$, the ratio of the trace anomaly divided by 4 over the radiation energy density
of the symmetric phase at the nucleation temperature, $\alpha$, and the speed of the phase transition at the nucleation temperature, $\beta$, as follows:
\begin{equation}
\frac{S_E(T_n)}{T_n} \simeq -\log\frac{(g_*\pi^2/30)^2 v_R^4}{9M_*^4}
\end{equation}
where the $g_*$ is the effective relativistic degrees of freedom and $M_*$ is the reduced Planck mass.
\begin{equation}
\alpha =  \frac{1}{\frac{\pi^2}{30}g_* T_n^4} \left(\frac{1}{-\frac{T}{4}\frac{\partial \Delta V}{\partial T} +\Delta V}\right)_{T=T_n},
\end{equation}
where $\Delta V = V_{\text{eff}}(0, T_n) - V_{\text{eff}}(v_n, T_n)$ is the potential difference between the two phases at the nucleation temperature.

\begin{equation}
\frac{\beta}{H} = T_n \frac{d}{dT}\left(\frac{S_E}{T}\right)\Bigg|_{T=T_n}.
\end{equation}

\subsection{Parameter Space Dependence}

To understand the gravitationanl waves from the phase transtions, we investigate how the phase transition parameters depend on the fundamental model parameters. We focus on the two most phenomenologically relevant couplings: the $SU(2)_R$ gauge coupling $g_R$ and the singlet-Higgs coupling $\lambda$.

\begin{figure}[htb]
\centering
\includegraphics[width=0.9\textwidth]{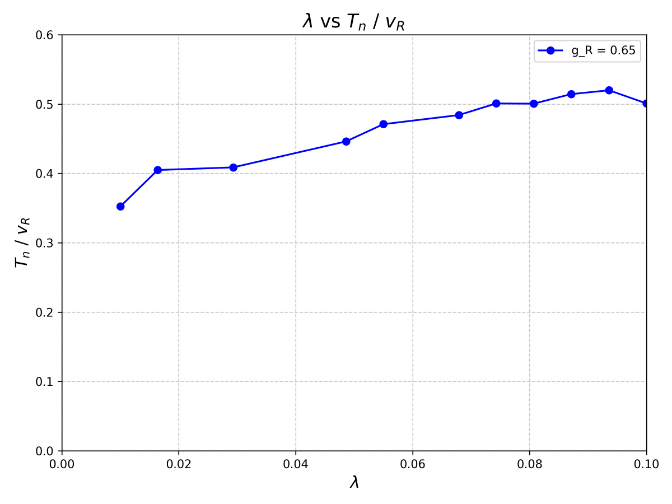}
\caption{Dependence of the nucleation temperature $T_n/v_R$ parameter on $\lambda$ for $g_R=g_L=0.655$. For other parameters, we fix $\xi = 5\times 10^8$ GeV, $k=0.01$, $m_H=10^3$ GeV, $Y_N=0.3$.}
\label{fig:gR_dependence}
\end{figure}

Fig.~\ref{fig:gR_dependence} shows the normalized nucleation temperature $T_n/v_R$ as a function of the singlet-Higgs coupling $\lambda$ for $g_R = 0.65$. The other parameters are fixed at $\xi = 5\times 10^8$ GeV, $k=0.01$, $m_H=10^3$ GeV, and $Y_N=0.3$. The ratio $T_n/v_R$ increases monotonically with $\lambda$, rising from $T_n/v_R \sim 0.36$ at $\lambda = 0.01$ to $T_n/v_R \sim 0.52$ at $\lambda = 0.10$. This behavior reflects the interplay between the singlet-Higgs coupling and the thermal barrier structure. 
For the phenomenologically motivated parity-symmetric case with $g_R = g_L = 0.65$ shown here, the range $\lambda \sim 0.01$--$0.05$ yields $T_n/v_R \sim 0.36$--$0.45$, corresponding to moderate supercooling that produces strong first-order transitions while maintaining peak frequencies $f \sim 0.1$--$1$ Hz within the target bands of future space-based gravitational wave observatories.

\begin{figure}[htb]
\centering
\includegraphics[width=0.46\textwidth]{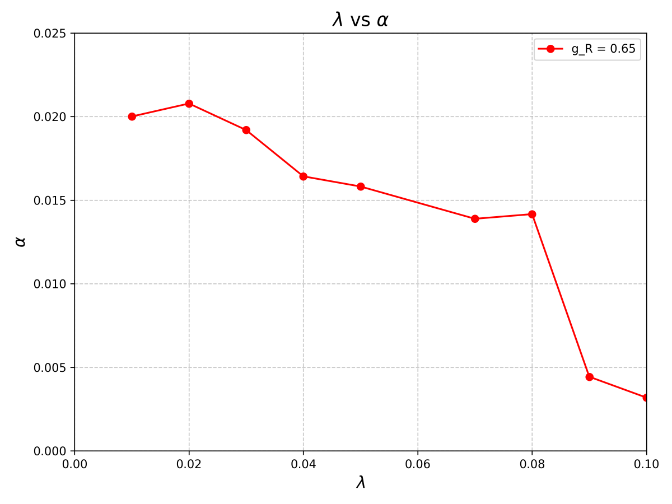}
\includegraphics[width=0.46\textwidth]{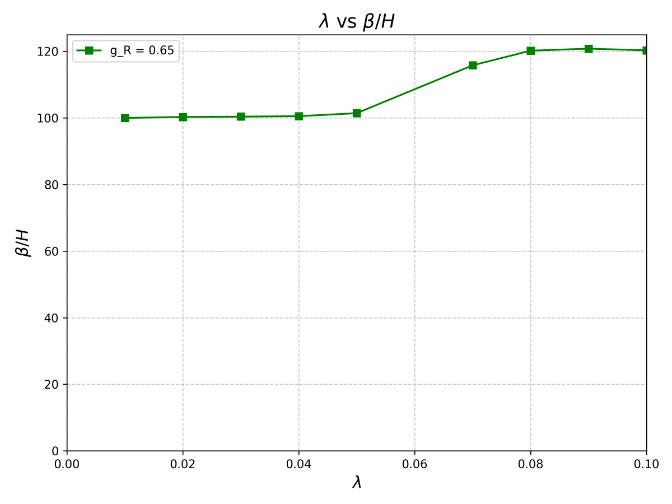}
\caption{Dependence of phase transition parameters $\alpha$ (Left) and $\beta/H$(Right) on $\lambda$$g_R=g_L=0.65$. For other parameters, we fix $\xi = 5\times 10^8$ GeV, $k=0.01$, $m_H=10^3$ GeV, $Y_N=0.3$.}
\label{fig:gR_dependence2}
\end{figure}

Figs.~\ref{fig:gR_dependence2} show the dependence of the phase transition parameters $\alpha$ (left panel) and $\beta/H$ (right panel) on the singlet-Higgs coupling $\lambda$ for the parity-symmetric case $g_R = g_L=0.65$. The other parameters are fixed at $\xi = 5\times 10^8$ GeV, $k=0.01$, $m_H=10^3$ GeV, and $Y_N=0.3$. The $\lambda$ dependencies of $\alpha$ and $\beta/H$ have important implications for gravitational wave phenomenology. Since the gravitational wave amplitude scales approximately as $h^2\Omega_{\rm GW} \propto \alpha^2/(\beta/H)$, the signal strength is dominated by the transition strength parameter. The weak $\beta/H$ variation means that the gravitational wave amplitude is primarily controlled by the dramatic $\lambda$ dependence of $\alpha$. For the phenomenologically interesting range $\lambda \sim 0.01$--$0.05$, we obtain $\alpha \sim 0.014$--$0.021$ and $\beta/H \sim 100$--$102$, yielding moderately strong first-order transitions capable of producing detectable gravitational waves.

\section{Gravitational Wave Spectrum}

\subsection{Production Mechanisms}

Gravitational waves arise from three sources:
the sound waves in the plasma, the magnetohydrodynamic turbulence 
, the bubble collisions \cite{Huber:2008hg}\cite{Binetruy:2012ze}\cite{Hindmarsh:2013xza}.

The sound waves are the dominant contributions, which are given by
\begin{equation}
\Omega_{\text{sw}}h^2 = 2.65 \times 10^{-6} \left(\frac{\beta/H}{100}\right)^{-1} \left(\frac{\kappa_{\text{sw}} \alpha}{1+\alpha}\right)^2 \left(\frac{g_*}{100}\right)^{-1/3} v_w S_{\text{sw}}(f),
\end{equation} 
with efficiency factor $\kappa_{\text{sw}} = \alpha/(0.73 + 0.083\sqrt{\alpha} + \alpha) \approx 0.10$ and spectral shape:
\begin{equation}
S_{\text{sw}}(f) = \left(\frac{f}{f_{\text{sw}}}\right)^3 \left(\frac{7}{4 + 3(f/f_{\text{sw}})^2}\right)^{7/2}
\end{equation}
The peak frequency is given by
\begin{equation}
f_{\text{sw}} = 1.9 \times 10^{-5} \left(\frac{T_n}{100 \text{ GeV}}\right) \left(\frac{g_*}{100}\right)^{1/6} \frac{1}{v_w} \text{ Hz}.
\end{equation}

The contribution from magnetohydrodynamic turbulence is given by:
\begin{equation}
\Omega_{\text{turb}}h^2 = 3.35 \times 10^{-4} \left(\frac{\beta/H}{100}\right)^{-1} \left(\frac{\kappa_{\text{turb}} \alpha}{1+\alpha}\right)^{3/2} \left(\frac{g_*}{100}\right)^{-1/3} v_w S_{\text{turb}}(f)
\end{equation}
where $\kappa_{\text{turb}} \approx 0.05$ is the efficiency factor for turbulence. The spectral shape is:
\begin{equation}
S_{\text{turb}}(f) = \frac{(f/f_{\text{turb}})^3}{[1 + (f/f_{\text{turb}})]^{11/3} (1 + 8\pi f/h_*)}
\end{equation}
with $h_* = 1.65 \times 10^{-5} \left(\frac{T_n}{100 \text{ GeV}}\right) \left(\frac{g_*}{100}\right)^{1/6}$ Hz. The peak frequency is:
\begin{equation}
f_{\text{turb}} = 2.7 \times 10^{-5} \left(\frac{T_n}{100 \text{ GeV}}\right) \left(\frac{g_*}{100}\right)^{1/6} \frac{1}{v_w} \text{ Hz}
\end{equation}

The contribution from bubble collisions is given by
\begin{equation}
\Omega_{\text{coll}}h^2 = 1.67 \times 10^{-5} \left(\frac{\beta/H}{100}\right)^{-2} \left(\frac{\kappa_{\text{coll}} \alpha}{1+\alpha}\right)^2 \left(\frac{g_*}{100}\right)^{-1/3} \left(\frac{0.11 v_w^3}{0.42 + v_w^2}\right) S_{\text{coll}}(f)
\end{equation}
where $\kappa_{\text{coll}} \approx 0.05$ is the efficiency factor for bubble collisions. The spectral shape is:
\begin{equation}
S_{\text{coll}}(f) = \frac{3.8 (f/f_{\text{coll}})^{2.8}}{1 + 2.8 (f/f_{\text{coll}})^{3.8}}
\end{equation}
with peak frequency:
\begin{equation}
f_{\text{coll}} = 1.65 \times 10^{-5} \left(\frac{T_n}{100 \text{ GeV}}\right) \left(\frac{g_*}{100}\right)^{1/6} \left(\frac{\beta}{H}\right) \frac{1}{v_w} \text{ Hz}
\end{equation}

The total gravitational wave spectrum is the sum of the three contributions:
\begin{equation}
\Omega_{\text{GW}}h^2 = \Omega_{\text{sw}}h^2 + \Omega_{\text{turb}}h^2 + \Omega_{\text{coll}}h^2
\end{equation}

The detectability of our predicted gravitational wave spectrum depends crucially on the sensitivity curves of future space-based and ground-based detectors \cite{LISACosmologyWorkingGroup:2022jok}. The Laser Interferometer Space Antenna (LISA) is expected to achieve a sensitivity of $h^2\Omega_{\text{GW}} \sim 10^{-12}$ in the frequency range $10^{-4}$ to $10^{-1}$ Hz, with optimal sensitivity around $10^{-2}$ Hz. The Big Bang Observer (BBO) and DECi-hertz Interferometer Gravitational wave Observatory (DECIGO) are proposed space-based missions with improved sensitivity, targeting $h^2\Omega_{\text{GW}} \sim 10^{-15}$-$10^{-17}$ in the $10^{-2}$ to $10^2$ Hz range. On the ground, the Einstein Telescope (ET) and Cosmic Explorer (CE) will extend the observational window to higher frequencies ($1$-$10^4$ Hz) with sensitivities reaching $h^2\Omega_{\text{GW}} \sim 10^{-9}$-$10^{-10}$ in their respective bands. Our predicted peak amplitude of $h^2\Omega_{\text{GW}} \sim 5 \times 10^{-11}$ at $f \sim 0.018$ Hz falls squarely within LISA's optimal sensitivity range, while also being accessible to BBO/DECIGO with higher signal-to-noise ratios.

\begin{figure}[htb]
\centering
\includegraphics[width=0.45\textwidth]{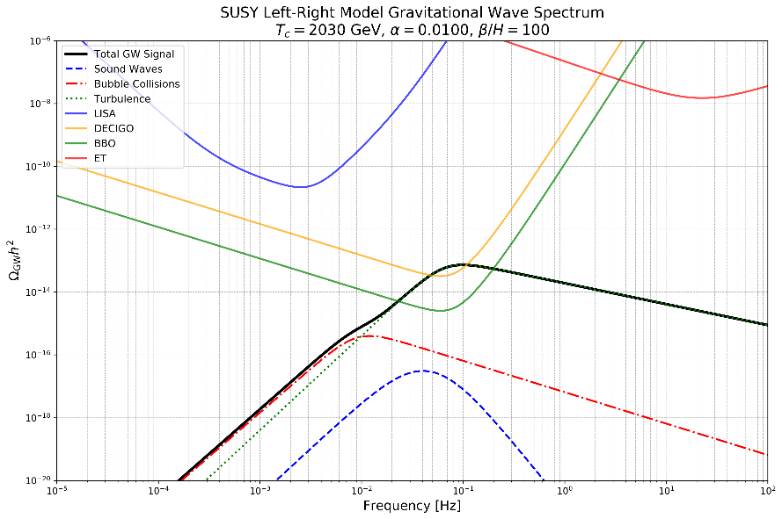}
\includegraphics[width=0.45\textwidth]{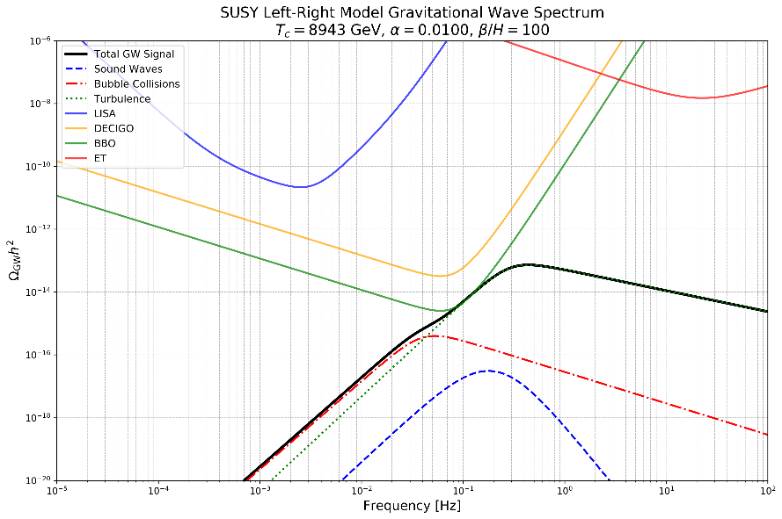}
\\
\includegraphics[width=0.45\textwidth]{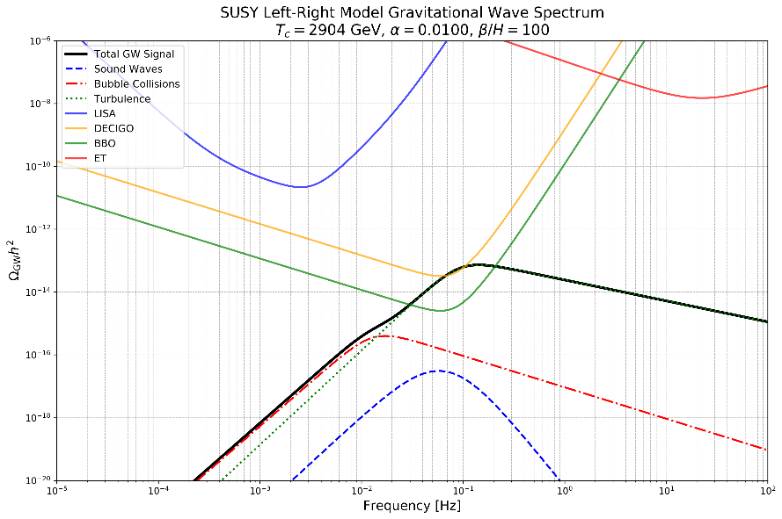}
\includegraphics[width=0.45\textwidth]{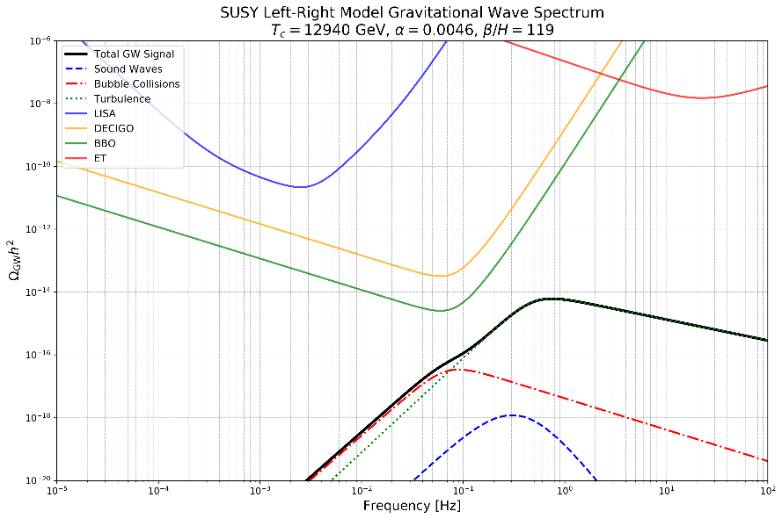}
\caption{Gravitational wave spectrum: The black solid lines are the total spectrum. The green dotted lines are the sound waves, the red dashed-dotted lines are the bubble collisions, the blue dashed lines are the turbulence contributions.
The LISA, DECIGO, BBO, ET sensitiviy lines are shown with the
blue solid line, the yellow solid line, the green solid line, and the red solid lines. All panels assume $gR=gL=0.65$.
(a).Left-upper Panel:  $\lambda=0.05$, $\xi=5\times 10^8$ GeV.
(b).Right-upper Panel: $\lambda=0.01$, $\xi= 10^8$ GeV.
(c).Left-lower Panel:  $\lambda=0.05$, $\xi=1.25\times 10^8$ GeV.
(d).Right-lower Panel: $\lambda=0.01$, $\xi=2.5\times 10^8$ GeV.
}
\label{fig:GWspectrum1}
\end{figure}

In Fig.\ref{fig:GWspectrum1}, the total gravitational wave energy density is shown as black solid lines, with individual contributions from sound waves (green dotted), bubble collisions (red dash-dotted), and magnetohydrodynamic turbulence (blue dashed). Sensitivity curves for planned detectors are overlaid: LISA (blue solid), DECIGO (yellow solid), BBO (green solid), and Einstein Telescope (red solid). To preserve parity symmetry, we set $g_R=g_L=0.65$ for the four cases, while varying two key parameters:  (a) $\lambda=0.05$, $\xi=5\times10^8$ GeV;
(b) $\lambda=0.01$, $\xi= 10^8$ GeV; 
(c) $\lambda=0.05$, $\xi=1.25\times10^8$ GeV;
(d) $\lambda=0.01$, $\xi=2.5\times10^7$ GeV. 
Sound waves dominate the signal across all parameter choices, confirming the model's robust gravitational wave predictions.
%For the case (d), the amplitude with the frequency $\sim 0.1$ Hz is mergenally overlap with the BBO sensitivity line.
In case (d),  the amplitude of the total gravitational wave signal at a frequency of approximately 0.1 Hz  overlaps with the sensitivity curve of the BBO detector. This suggests that if the universe underwent a phase transition with parameters similar to case (d), the BBO mission would have a chance of detecting it \footnote{If we relax the parity-symmetric assumption $g_R=g_L$ and instead consider a larger right-handed gauge coupling $g_R\sim 1$ (while keeping $g_L$ at its Standard Model value), we found that there is a prameter region where the total gravitational wave amplitude is significantly enhanced since  a larger $g_R$ strengthens the phase transition. The enhanced gravitational wave signal shifts to frequencies and amplitudes that overlap with the DECIGO sensitivity curve.
}.

\section{Conclusions}

We have considered that gravitational waves from the phase transition in the supersymmetric left-right model where the strong CP problem resolved. This comprehensive analysis of phase transition dynamics in the supersymmetric left-right symmetric model shows compelling prospects for gravitational wave detection and fundamental physics exploration. Our calculations demonstrate that the $SU(2)_R \times U(1)_{B-L} \to U(1)_Y$ symmetry breaking transition exhibits the characteristics necessary for observable gravitational wave production.
We found  the parameter region where the peak frequency $f \sim 0.1-1$ Hz lies  overlapping with the DECIGO/BBO sensitivity curves. Future missions may achieve even higher sensitivity. The gravitational wave amplitude scales favorably with both the $SU(2)_R$ gauge coupling $g_R$ and the singlet-Higgs coupling $\lambda$, while maintaining peak frequencies accessible to planned detectors.

We have demonstrated that well-motivated theoretical solutions to fundamental problems can yield specific, testable predictions for next-generation gravitational-wave detectors, establishing gravitational-wave astronomy as a powerful tool for probing physics beyond the Standard Model. The convergence of theoretical motivation, observational capability, and a clear experimental timeline makes this supersymmetric left-right model a particularly promising candidate for discovery in the coming decade of gravitational-wave science.

This provides a remarkable connection between a theoretical solution to the strong CP problem and potentially observable signatures in gravitational wave astronomy. The detection of such a signal would provide unique insight into physics at energy scales far beyond the reach of collider experiments.

\section*{Acknowledgments}

This work is partially supported by Scientific Grants by the Ministry of Education, Culture, Sports, Science and Technology of Japan, No. 23H03392 (N.H.) and No. 19K147101 (T.Y.).

\bibliographystyle{apsrev4-1}

\end{document}